\documentstyle[referee,psfig,epsf]{l-aa}
\def\lsim{\lower.5ex\hbox{$\; \buildrel < \over \sim \;$}}
\def\gsim{\lower.5ex\hbox{$\; \buildrel > \over \sim \;$}}
\newcommand{\eqb}{\begin{eqnarray}}
\newcommand{\eqe}{\end{eqnarray}}

\newbox\grsign \setbox\grsign=\hbox{$>$} \newdimen\grdimen \grdimen=\ht\grsign
\newbox\simlessbox \newbox\simgreatbox
\setbox\simgreatbox=\hbox{\raise.5ex\hbox{$>$}\llap
   {\lower.5ex\hbox{$\sim$}}}\ht1=\grdimen\dp1=0pt
\setbox\simlessbox=\hbox{\raise.5ex\hbox{$<$}\llap
   {\lower.5ex\hbox{$\sim$}}}\ht2=\grdimen\dp2=0pt
\def\gsim{\mathrel{\copy\simgreatbox}}
\def\lsim{\mathrel{\copy\simlessbox}}
 
\def\ref{\par\noindent\hangindent=1.5cm}
\begin{document}
\thesaurus{06(02.01.2, 02.02.1, 08.14.1, 02.08.1)}
\title{Neutron Disks Around Black Holes}
\author{Sandip K. Chakrabarti and Banibrata Mukhopadhyay}
\institute{S.N. Bose National Centre for Basic Sciences, JD Block, Salt Lake, Calcutta 
700091, INDIA}

\offprints{S.K. Chakrabarti}
\date{Received 6 October 1998 ; accepted 13 January 1999\\
\vskip0.05cm
Astron. Astrophys. 344, 105 - 110 (1999)}
\maketitle
\markboth{Neutron Disks}{}

\maketitle

\begin{abstract}

We study nucleosynthesis in low accretion rate hot advective flows
around black holes. We find that matter is generally 
photo-dissociated into protons and neutrons inside the disk. 
These neutrons stay around black holes for longer time
because they are not coupled to magnetic fields while the protons
accrete into the hole. We find the nature of the resulting 
neutron disks and estimate the rate at which these disks 
contaminate the surroundings.

 \end{abstract}

\section{Introduction}

Angular momentum in accretion disks around black holes 
must deviate from a Keplerian distribution, since the
presence of ion, radiation or inertial pressure gradient forces
become as significant as the gravitational and 
centrifugal forces (see Chakrabarti 1996a; Chakrabarti 
1996b and references therein). The inertial 
pressure close to a black hole is high, because, 
on the horizon, the inflow velocity must be 
equal to the velocity of light. For causality, the 
velocity of sound must be less than the velocity of light.
In fact, in the extreme equation of state of
$P=\frac{c^2}{3} \rho$ (where $c$ is the velocity of light
and $P$ and $\rho$ are the isotropic pressure and mass density
respectively), the sound speed is only $c/\sqrt{3}$. Thus, the flow must 
pass through a sonic point and become supersonic before entering into the 
horizon. A flow which must pass through a sonic point must also be 
sub-Keplerian (Chakrabarti 1996b and references therein), and this causes the deviation.
If the accretion rate is low, the flow cools down only by 
inefficient bremsstrahlung and Comptonization processes, unless the magnetic 
field is very high (Shvartsman 1971; Rees 1984; Bisnovatyi-Kogan 1998). 
This hot flow can undergo significant nucleosynthesis 
depending on the inflow parameters. Earlier, in the context of 
thick accretion disks calculations of changes in composition inside
an accretion disk were carried out 
(Chakrabarti et al. 1987; Hogan \& Applegate 1987; 
Arai \& Hashimoto 1992; Hashimoto et al. 1993), but 
the disk models used were not completely self-consistent, in that neither 
the radial motion, nor the cooling and heating processes were included
fully self-consistently. Secondly, only high accretion rates were used.
As a result, the viscosity parameter required for a significant nuclear burning
was extremely low ($\alpha_{vis} < 10^{-4}$).
In the present paper, we do the computation after including the radial
velocity in the disk and the heating and cooling processes. We largely follow 
the solutions of Chakrabarti (1996b) to obtain the thermodynamic conditions
along a flow.
 
Close to a black hole horizon, the viscous time scale is so large compared 
to the infall time scale that the specific angular momentum $\lambda$ 
of matter remains almost constant and sub-Keplerian independent of viscosity 
(Chakrabarti 1996a,b; Chakrabarti 1989). 
Because of this, as matter accretes, the centrifugal force $\lambda^2/x^3$ 
increases much 
faster compared to the gravitational force $GM/x^2$ (where $G$ and $M$ are the 
gravitational constant and the mass of the black hole respectively, 
$\lambda$ and $x$ are the dimensionless angular momentum and the 
radial distance from the black hole).
As a result, close to the black hole (at $x \sim \lambda^2/GM$) 
matter may even virtually stop to form standing shocks (Chakrabarti 1989). 
Shock or no-shock, as the flow slows down,
the kinetic energy of matter is converted
into thermal energy in the region where the centrifugal force dominates.
Hard X-rays and $\gamma$-rays are expected from here
(Chakrabarti \& Titarchuk, 1995). In this centrifugal 
pressure supported hot `boundary layer' (CENBOL) of the black hole
(Chakrabarti et al. 1996) we find that
for low accretion rates, $^4He$ of the infalling 
matter is completely photo-dissociated and no $^7Li$ could be produced. 
In this region,  about ten to twelve percent of matter is found to be made 
up of pure neutrons. These neutrons should not accrete very fast because 
of very low magnetic viscosity associated with neutral particles 
(Rees et al. 1982)  while protons are dragged towards the
central black hole along with the field lines. Of course, both 
the neutrons and protons would have `normal' ionic viscosity, 
and some slow accretion of protons (including 
those produced after neutron decay) would still be possible. 
In contrast to neutron stars, the {\it neutron disks} which we
find are not dense. Nevertheless, they can participate in the formation of 
neutron rich isotopes and some amount of deuterium. They can 
be eventually dispersed into the galaxy through jets and outflows, 
which come out of CENBOL (Chakrabarti 1998; Das \& Chakrabarti 1998)
thereby possibly influencing the metallicity of the galaxy.

On the equatorial plane, where the viscosity is the highest,
a Keplerian disk deviates to become sub-Keplerian {\it very close} 
to the black hole (Chakrabarti \& Titarchuk 1995; Wiita 1982).
Away from the equatorial plane, viscosity is lower and the flow deviates
from a Keplerian disk farther out. This is because the angular 
momentum transport is achieved by viscous stresses. Weaker the 
viscosity, longer is the distance through which angular momentum 
goes to match with a Keplerian disk. When the viscosity of the disk 
is decreased on the whole, the Keplerian disk recedes 
from the black hole forming quiescence states
when the objects become very faint in X-rays (Ebisawa et al. 1996). Soft photons from the Keplerian disk are 
intercepted by this sub-Keplerian boundary layer (CENBOL) and photons 
are energized through Compton scattering process. 
For higher Keplerian rates, electrons and protons cool
down completely and the black hole is in a {\it soft state} 
(Tanaka \& Lewin 1995). Here, bulk motion Comptonization 
produces the power-law tail of slope $\alpha
\sim 1.5$ (Chakrabarti \& Titarchuk 1995; Titarchuk et al. 1997). For lower Keplerian rates, 
the Compton cooling is incomplete and the 
temperature of the boundary layer remains close to the virial value, 
$$
T_p \sim \frac{1}{2 k} m_p c^2 \frac{x_g}{x} = 5.2 \times 10^{11} \left( \frac{10}{x/x_g} 
\right) \ \ {\rm ^o K}.
\eqno{(1)}
$$
In this case, bremsstrahlung is also important and the black hole is said 
to be in a {\it hard state} with energy spectral index $\alpha$ 
($F_\nu \sim \nu^{-\alpha}$, where $\nu$ is the 
frequency of the photon) close to $0.5$. In Eq. (1), $m_p$ is the mass of 
the proton, $x_g=2GM/c^2$ is the Schwarzschild radius of the 
black hole, and $c$ is the velocity of light. (In future, we 
measure the distances and velocities in units of $x_g$ and $c$.) 
In this low Keplerian rate, electrons are cooler typically by a factor 
of $(m_p/m_e)^{1/2}$ unless the magnetic field is very high. 
Present high energy observations seem to support the 
apparently intriguing aspects of black hole accretion mentioned
above. For instance, the constancy of (separate) spectral slopes in soft and
hard states has been observed by many (Ebisawa et al. 1994; Miyamoto et al. 1991; Ramos 
et al. 1997; 
Grove et al. 1998; Vargas et al. 1997). ASCA observations of Cygnus X-1 seem to indicate 
that the 
inner edge of the Keplerian component is located at around $15R_g$ (instead of $3R_g$)
(Gilfanov et al. 1997).   HST FOS observations of the black hole
candidate A0620-00 in quiescent state seems to have very faint Keplerian features 
(McClintock et al. 1995) indicating the Keplerian component to be farther out at low accretion 
rates. 
Bulk motion Comptonization close to the
horizon has been considered to be a possible cause of the power-law tail in
very soft states (Crary et al., 1996; Ling et al. 1997; Cui et al. 1997).
However, some alternative modes may not be ruled out to explain some of these features.

This observed and predicted dichotomy of states of black hole spectra 
motivated us to investigate the nuclear reactions thoroughly for both 
the states, but we report here the results obtained in the more important 
case, namely, when the flow is hotter, i.e., for hard states. 
We use $255$ nuclear elements in the thermo-nuclear network starting 
from protons, neutrons, deuterium etc. till $^{72}Ge$ and the nuclear
reaction rates valid for high temperatures. We assume that 
accretion on the galactic black hole is taking place from a disk where 
matter is supplied from a normal main sequence star. That is, 
we choose the abundance of the injected matter to be that of the sun. Because 
of very high temperature, the result is nearly independent of the 
initial composition, as long as reasonable choices are made. When
accretion rates are higher, the advective region becomes cooler
and  very little nucleosynthesis takes place, the results are
presented elsewhere (Mukhopadhyay 1998; Mukhopadhyay \& Chakrabarti
1998).

As hot matter approaches a black hole, photons originated by the bremsstrahlung
process, as well as those intercepted from the Keplerian disk,
start to photo-dissociate deuterium and helium in the advective
region. There are two challenging issues at this stage which 
we address first: (a) Thermodynamic 
quantities such as density and temperature inside a disk are 
computed using a {\it thin disk} approximation, i.e., the 
vertical height $h(x)$ at a radial distance $x$ very small
compared to $x$ ($h(x) <<x$), and assuming
the flow to be instantaneously in vertical equilibrium.
However, at a low rate, it is easy to show that the disk is 
optically thin in the vertical direction $\int_0^{h(x)} 
{\rho \sigma dh} <1$ ($\sigma$ is the Thomson scattering
cross-section). However, soft photons from the Keplerian 
disk enter radially and $\int_1^{x_s} \rho \sigma dx >1$, generally.
In fact, this latter possibility changes the soft photons 
of a few keV from a Keplerian disk to energies up to $ \sim 1$MeV 
by repeated Compton scattering (Sunyaev \& Titarchuk 1980; 
Chakrabarti \& Titarchuk 1995) while keeping the photon number 
strictly constant. The spectrum of the resultant photons 
emitted to distant observers becomes a power law $F_\nu \sim \nu^{-\alpha}$ 
instead of a blackbody, where $\alpha \sim 0.5$ for hard state
and $\alpha \sim 1.5$ for soft states of a black hole. (b) Now that 
the spectrum is not a blackbody, strictly speaking, the 
computation of photo-disintegration  rate that is standard in the 
literature (which utilizes a Planckian spectrum) cannot be followed. 
Fortunately, this may not pose a major problem. As we shall 
show, the standard photo-disintegration rate yields
a lower limit of the actual rate that takes place in the presence 
of power-law photon spectra. Thus, usage of the correct
rate obtainable from a power-law spectrum would, 
if anything, strengthen our assertion
about the photo-disintegration around a black hole.
After photo-disintegration by these hard photons,
all that are left are protons and neutrons. The exact location where
the dissociation actually starts may depend on the detailed photon spectrum,
i.e., optical depth of this boundary layer and the electron temperature. 

The plan of the present paper is the following: in the next section, we 
present briefly the hydrodynamical model using which the thermodynamic 
quantities such as the density and temperature inside the inner accretion 
disk are computed. We also present the model parameters we employ. 
In Sect. 3, we present results of nucleosynthesis inside a disk. 
Finally, in Sect. 4, we present out concluding remarks.

\section{Model Determining the Thermodynamic Conditions}

We chose the units of distance, time and mass to be $2GM/c^2$, $2GM/c^3$ and $M$
where, $G$ is the gravitational constant, $M$ is the mass of the black hole,
and $c$ is the velocity of light. To keep the problem tractable without 
sacrificing the salient features, we use a well understood model of the 
accretion flow close to the black hole. We solve the following equations 
(Chakrabarti 1996a,b) to obtain the thermodynamic quantities:

\noindent (a) The radial momentum equation:

$$
\vartheta \frac{d\vartheta}{dx} +\frac{1}{\rho}\frac{dP}{dx}
+\frac {\lambda_{Kep}^2-\lambda^2}{x^3}=0,
\eqno{(2a)}
$$

\noindent (b) The continuity equation:

$$
\frac{d}{dx} (\Sigma x \vartheta) =0 ,
\eqno{(2b)}
$$

\noindent (c) The azimuthal momentum equation:

$$
\vartheta\frac{d \lambda(x)}{dx} -\frac {1}{\Sigma x}\frac{d}{dx}
(x^2 W_{x\phi}) =0 ,
\eqno{(2c)}
$$

\noindent (d) The entropy equation:

$$
\Sigma v T \frac{ds}{dx} = \frac{h(x) \vartheta}{\Gamma_3 - 1}(\frac{dp}{dx} -
\Gamma_1 \frac{p}{\rho}) = Q^+_{mag}+Q^+_{nuc}+Q^+_{vis}-Q^-  
= Q^+ - g(x, {\dot m}) Q^+ = f(\alpha, x, {\dot m}) Q^+ .
\eqno{(2d)}
$$
Here, $Q^+$ and $Q^-$ are the heat gained and lost by the flow, and ${\dot m}$ 
is the mass accretion rate in units of the Eddington rate. Here, we have 
included the possibility of magnetic heating $Q^+_{mag}$ 
(due to stochastic fields; Shvartsman 1971; Shapiro, 1973; Bisnovatyi-Kogan, 1998)
and nuclear energy release $Q^+_{nuc}$ as well (cf. Taam \& Fryxall 1985)
while the cooling is provided by bremsstrahlung, 
Comptonization, and endothermic reactions and neutrino emissions.
A strong magnetic heating might equalize ion and electron
temperatures (e.g. Bisnovatyi-Kogan 1998) but this would not
affect our conclusions. On the right hand side, we wrote $Q^+$ collectively 
proportional to the cooling term for simplicity (purely on 
dimensional grounds). We use the standard definitions 
of $\Gamma$ (Cox \& Giuli 1968),
$$
\Gamma_3=1+\frac{\Gamma_1-\beta}{4-3\beta},
$$
$$
\Gamma_1=\beta + \frac{(4-3\beta)^2 (\gamma -1 )}{\beta + 12 (\gamma -1)(1-\beta)}
$$
and $\beta (x) $ is the ratio of gas pressure to total pressure,
$$
\beta(x) = \frac {\rho k T/\mu m_p}{\rho k T/\mu m_p + {\bar a} T^4/3 + B(x)^2/4\pi}
$$
Here, ${\bar a}$ is the Stefan constant, $k$ is the Boltzman constant, $m_p$
is the mass of the proton, $\mu$ is the mean molecular weight. 
Using the above definitions, Eq. (2d) becomes,
$$
\frac{4-3\beta}{\Gamma_1-\beta} [\frac{1}{T}\frac{dT}{dx}
-\frac{1}{\beta}\frac{ d \beta}{dx} -
\frac{\Gamma_1 - 1}{\rho}\frac{d\rho}{dx} ]
 = f(\alpha, x, {\dot m}) Q^+.
\eqno{(2e)}
$$
In this paper, we shall concentrate on solutions with constant $\beta$. 
Actually, we study in detail only the special cases, 
$\beta=0$ and $\beta=1$, so we shall liberally use $\Gamma_1=\gamma=\Gamma_3$.
We note here that unlike {\it self-gravitating} stars where $\beta=0$ 
causes instability, here this is not a problem. Similarly,
we shall consider the case for $f(\alpha, x, {\dot m})$ = constant, though
as is clear, $f\sim 0$ in the Keplerian disk region and 
probably much greater than $0$ near the black hole depending on the efficiency of
cooling (governed by ${\dot m}$, for instance). 
We use the Paczy\'nski-Wiita (1980) potential to describe the
black hole geometry. Thus, $\lambda_{Kep}$, the Keplerian angular
momentum is given by, $\lambda_{Kep}^2=x^3/2(x-1)^2$, 
exactly same as in general relativity.
$W_{x\phi}$ is the vertically integrated viscous stress,
$h(x)$ is the half-thickness of the disk at radial
distance $x$ (both measured in units of $2GM/c^2$)
obtained from vertical equilibrium assumption (Chakrabarti 1989)
$\lambda(x)$ is the specific angular momentum,
$\vartheta$ is the radial velocity, $s$ is the entropy density
of the flow. The constant $\alpha$ above is the Shakura-Sunyaev (1973)
viscosity parameter modified to include the pressure due to radial motion
($\Pi=W+\Sigma \vartheta^2$, 
where $W$ and $\Sigma$ are the integrated pressure and density
respectively; see Chakrabarti \& Molteni (1995) in the viscous stress.
With this choice, $W_{x\phi}$ keeps the specific angular 
momentum continuous across of the shock.

For a complete run, we supply the basic parameters, namely, the location of 
the sonic point through which flow must pass just outside the horizon 
$X_{out}$, the specific angular momentum at the inner edge of the 
flow $\lambda_{in}$, the polytropic index $\gamma$, the ratio $f$ of advected 
heat flux $Q_+-Q_-$ to heat generation rate $Q^+$, the viscosity parameter 
$\alpha_{vis}$ and the accretion rate ${\dot m}$. The derived quantities 
are: $x_{tr}$ where the Keplerian flow deviates to become sub-Keplerian, 
the ion temperature $T_p$, the flow density $\rho$, the radial velocity 
$v_r$ and the azimuthal 
velocity $\lambda/x$ of the entire flow from $x_{tr}$ to the
horizon. Temperature of the ions obtained from above equations 
is further corrected using a cooling factor $F_{Comp}$
obtained from the results of radiative transfer of
Chakrabarti \& Titarchuk (1995).
Electrons cool due to Comptonization, but they cause the ion
cooling also since ions and electrons are coupled by Coulomb interaction.
$F_{Comp}$, chosen here to be constant in the advective region, 
is the ratio of the ion temperature computed from hydrodynamic 
(Chakrabarti 1996b) and radiation-hydrodynamic (Chakrabarti \& 
Titarchuk 1995) considerations.

\section{Results of Nucleosynthesis Calculations}

In the first example, we start with a relativistic flow (polytropic 
index $\gamma=4/3$) with the accretion rate ${\dot M}=0.01 {\dot M}_{Edd}$, 
where, ${\dot M}_{Edd}$ 
is the Eddington accretion rate. We use the mass of the central black hole 
to be $M=10M_\odot$ throughout. We choose a very high viscosity and the 
corresponding $\alpha$ parameter (Shakura \& Sunyaev 1973) is $0.2$ in the 
sub-Keplerian regime. The cooling is not as efficient as in a Keplerian disk: 
$Q^- \sim 0.9 Q^+$, where, $Q^+$ and $Q^-$ are the heat generation and heat 
loss rates respectively. The specific angular momentum at the inner edge 
is $\lambda_{in}=1.65$ (in units of $2GM/c$). The flow deviates from 
a Keplerian disk at $4.15$ Schwarzschild radii. It is to be noted 
that $Q^-$ includes {\it all possible} types of cooling, such 
as bremsstrahlung, Comptonization as well as cooling due to neutrino 
emissions. We assume that the flow is magnetized so that only ions have 
larger viscosity. Due to poor supply of the soft photons from 
Keplerian disks, the Comptonization in the boundary layer is 
not complete: we assume a standard value (Chakrabarti, \& Titarchuk 1995)
in this regime: $F_{Comp} \sim 0.1$, i.e., ions (in te radiation-hydrodynamic 
solution) are one-tenth as hot as obtained from the hydrodynamic solutions. 
[For high accretion rate, ${\dot m}\gsim 0.3$, 
$F_{Comp} \sim 0.001$ and ions and electrons both 
cool to a few KeV ($\sim 10^7$ $^o$K)]. 
The typical density and temperature near the marginally stable orbit are
$\rho_{x=3} \sim 8.5 \times 10^{-8}$ gm cm$^{-3}$ and $7.5 \times 10^9$ $^o$K
respectively where the thermonuclear depletion rates $N_A <\sigma v>$ for the
$D \rightarrow p + n$, $^4He \rightarrow  D + D$ and $^4He +^4He = ^7Li +p$
reactions are given by $ 1.6 \times 10^{14}$ gm$^{-1}$ s$^{-1}$, $4 \times 10^{-3}$ 
gm$^{-1}$ s$^{-1}$
and $ 1.9 \times 10^{-12}$ gm$^{-1}$ s$^{-1}$ respectively. 
Here, $N_A$ is the element abundance on the
left, $\sigma$ is the reaction cross-section, $v$ 
is the Maxwellian average velocity of the reactants. At these 
rates, the time scales of these reactions are given by,
$4 \times 10^5$s, $5 \times 10^{11}$s and $4 \times 10^{20}$s respectively
indicating that the deuterium burning is the fastest of the reactions. 
In fact, it would take about a second to burn initial deuterium 
with $Y_D=10^{-5}$. The $^7Li$ does not form at all because 
the $^4He$ dissociates to $D$ much faster.

The above depletion rates have been computed assuming Planckian photon distribution 
corresponding to ion temperature $T_p$. The wavelength $\lambda_{Planck}$ at which 
the brightness is highest at $T=T_p$ is shown in Fig. 1 in 
the dashed curve (in units of $10^{-11}$ cm). Also shown is the {\it average} 
wavelength of the photon $\lambda_{Compton}$ (solid curve) obtained from the 
spectrum $F_\nu \sim \nu^{-\alpha}$. The average has been performed over the
region $2$ to $50$keV of the photon energy in which the hard component is 
usually  observed
$$
< F_\nu > =\frac{\int_{\nu_{min}}^{\nu_{max}} F_\nu d\nu}{\int_{\nu_{min}}^{\nu^{max}} 
d\nu} = \nu_{Compton}^{-\alpha}
\eqno{(3)}
$$
where, $\nu_{min}$ and $\nu_{max}$ are computed from $2$ and $50$keV respectively. 
The average becomes a function of the energy spectral index $\alpha$ 
($F_\nu \propto \nu^{-\alpha}$), which in turn depends on the ion and 
electron temperatures of the medium. We follow Chakrabarti
\& Titarchuk (1995) to compute these relations. We note that
$\lambda_{Compton}$ is {\it lower} compared to $\lambda_{Planck}$ for all ion
temperatures we are interested in. Thus, the disintegration
rate with Planckian distribution that we employed in this 
paper is clearly a lower limit. Our assertion of the formation of 
a neutron disk should be strengthened when Comptonization 
is included.

Figure 2 shows the result of the numerical simulation for the disk model 
mentioned above. Logarithmic abundance of neutron $Y_n$ is plotted against the 
logarithmic distance from the black hole. First simulation produced the 
dash-dotted curve for the neutron distribution,
forming a miniature neutron torus. As fresh matter is added to the existing 
neutron disk, neutron abundance is increased as neutrons do not fall in
rapidly. Thus the simulation is
repeated several times in order to achieve a converging steady 
pattern of the neutron disk. Although fresh neutrons are
deposited, the stability of the distribution is achieved through neutron 
decay and neutron capture reactions. Results after every ten iterations
are plotted. The equilibrium neutron torus remains around the black 
hole indefinitely. The neutron abundance is clearly very significant 
(more than ten per cent!). 

We study yet another case where the accretion rate is smaller (${\dot m}=0.001$) and 
the viscosity is so small ($\alpha=0.01$) and the disk so hot that the sub-Keplerian 
flow deviates from a Keplerian disk farther away at $x=85.1$. The polytropic index is 
that of a mono-atomic (ionized) hot gas $\gamma=5/3$. The Compton cooling factor 
is as above since it is independent of the accretion rates as long as the rate is 
low (Sunyaev \& Titarchuk 1980; Chakrabarti \& Titarchuk 1995). 
The cooling is assumed to be very inefficient because of 
lower density: $Q^- \sim 0.4 Q^+$. The specific angular 
momentum at the inner edge of the disk is $\lambda_{in}=1.55$.
In Fig. 3, we show the logarithmic abundances of proton (p), helium ($^4He$) 
and neutron (n) as functions of the logarithmic distance from the black hole. 
Note that $^4He$ dissociates completely at a distance of around $x=30$ where 
the density and temperatures are $\rho = 2.29 \times 10^{-11}$ gm cm$^{-3}$ 
and $T=6.3 \times 10^9$ $^o$K. Maximum temperature attained in this case 
is $T_{max} = 3.7 \times 10^{10}$ $^o$K. Both the neutrons and protons are 
enhanced for $x\lsim 30$, the boundary layer of the black hole. 
This neutron disk also remains stable despite neutron decay, 
since new matter moves in to maintain equilibrium. The $^7Li$ abundance is 
insignificant.

\section {Concluding Remarks}

In this paper, we have shown that hot flows may produce neutron disks
around black holes, where neutron abundance is 
significant. However, unlike neutron 
stars, the formation of which is accompanied by the production of neutron rich
isotopes, neutron disks do not produce significant neutron rich elements. 
Some fragile elements, such as deuterium, could be produced in the cooler 
outflows as follows:

Neutrons and protons may be released in space through winds which are produced
in the centrifugal barrier. These winds are common in black hole 
sources and earlier they have been attributed to the dispersal of magnetic 
fields to the galactic medium (Daly \& Loeb 1990; Chakrabarti et al. 1994). Recently, Chakrabarti (1998) and Das \& Chakrabarti (1998),
through a first ever self-consistent calculation of outflows out of accretion,
found that significant winds can be produced and for low enough 
accretion rates, disks may even be almost evacuated causing the formation of 
quiescence and inactive states  such as what is observed in V404 Cyg and our 
Galactic centre. If the temperature of the wind falls 
off as $1/z$ and density as $z^{-3/2}$ (as is expected from an outflow of 
insignificant rotation), the deuterium synthesis rate $n + p \rightarrow D$, 
increases much faster very rapidly than the reverse ($D \rightarrow n+p$) 
process. For instance, with density and 
temperature mentioned as in the earlier section, at $z=30 x_g$,
the forward rate ($N_A <\sigma v>$) is $0.12\times 10^{-5}$ while the reverse 
rate is much higher: $6.7 \times 10^{13}$. This results in the dissociation of
deuterium. However, at $z=300 x_g$, the above rates are $1.8 \times 10^{-8}$ and
$9.6 \times 10^{-6}$ respectively and at $z=3000 x_g$, the above rates are $1.3 \times 
10^{-8}$ and
$\sim 10^{-165}$ respectively. Thus a significant deuterium could be produced
farther out, say, starting from a distance of $\sim 10^3 x_g$. 
Ramadurai \& Rees (1985) suggested deuterium formation on 
the surface of ion tori. As we establish here, this process may be
feasible, only if these tori are vertically {\it very thick}: 
$z(x) \sim 10^3 x_g$. In any case, deuterium would be expected to form 
in winds and disperse.

\begin {figure}
\noindent {\small {\bf Fig. 1.}  Comparison of wavelength $\lambda_{Planck}$ at peak 
blackbody intensity
(dotted) with the mean (taken between $2$ and $50$keV) wavelength of the Comptonized power 
law spectrum 
(solid) of the emitted X-rays. Wavelengths are measured in units of $10^{-11}$cm.}
\end{figure}

\begin{figure}
\noindent {\small {\bf Fig. 2.} Formation of a steady {\it neutron torus} in a hot inflow. 
Intermediate iteration results (from bottom to top: 1st, 11th, 21st, 31st and 41st 
iterations respectively) of
the logarithmic neutron abundance $Y_n$ in the flow as a function of the logarithmic radial
distance ($x$ in units of Schwarzschild radius) are shown.  }
\end{figure}

\begin {figure}
\noindent {\small {\bf Fig. 3.} Variation of matter abundance $Y_i$ in
logarithmic scale  in a hot flow around a galactic black hole. Entire $^4He$ is
photodissociated at around $x=30x_g$ and the steady neutron disk is produced for $x<30$
which is not accreted.}
\end{figure}

In a typical case of a disk with an accretion rate of ${\dot M} \sim {\dot M}_{Edd}$,
the temperature is lower, but the density is higher. In that case, the photo-dissociation 
of $^4He$ is insignificant and typically the change in abundances of some of 
the elements, such as $^{16}O$, $^{20}Ne$ etc. could be around $\Delta Y \sim 
10^{-3}$ not as high as that of the neutron as in above cases where $\Delta 
Y_n \sim 0.1$. One could estimate the contamination of the galactic metalicity
due to nuclear reactions as we do for realistic models. Assume that, on an average, 
all the $N$ stellar black holes of equal mass $M$ have a non-dimensional 
accretion rate of around ${\dot m} \sim 1$ (${\dot m}={\dot M}/{\dot M}_{Edd}$). 
Let  $\Delta Y_i$ be the typical change in composition of this matter 
during the run and let $f_w$ be the fraction of the incoming flow that goes out as winds 
and outflows, then in the lifetime of a galaxy (say, $10^{10}$yrs), the total `change' 
in abundance of a particular species deposited to the surroundings by all the stellar 
black holes is given by:
$$
<\Delta Y_i> \cong 10^{-9} (\frac{\dot m}{1}) (\frac{N}{10^6}) (\frac{\Delta Y_i}{10^{-3}}) 
(\frac{f_w}{0.1} ) (\frac {M}{10 M_\odot}) (\frac{T_{gal}}{10^{10}}) (\frac{M_{gal}}{10^{13} M_\odot})^{-1}  .
\eqno{(4)}
$$
We here assume a conservative estimate that there are $10^6$ such stellar black holes 
(there number varies from $10^8$ (van den Heuvel 1992, 1998) to several thousands (Romani, 1998)
depending on assumptions made) and the mass of the host galaxy is around $10^{13}M_\odot$ 
and the lifetime of 
the galaxy during which such reactions are going on is about 
$10^{10}$Yrs. We believe that $<\Delta Y_i> \sim 10^{-9}$ is quite
reasonable for a typical case  when $\Delta Y_i \sim 10^{-3}$ 
and a fraction of ten percent of matter is blown off as winds. 
When $\Delta Y_i \sim 0.1$ or the outflow rate is higher 
(particularly in presence of strong centrifugal barrier) the 
contamination would be even higher.

It is to be noted that our assertion of formation of neutron disks 
around a black hole for very low accretion rate ${\dot M} \sim 
0.001-0.01 {\dot M}_{Edd}$ is different from that of the earlier 
results (Hogan, \& Applegate, 1987) where ${\dot M} \sim 10 {\dot M}_{Edd}$ was believed 
to be the more favourable accretion rate. This is because in last
decades the emphasis was on super-Eddington thick accretion tori. 
More recent computations suggest that advective regions are not as hot 
when the rates are very high. Another assertion of our work is 
that $^7Li$ should not be produced in accretion disks at all. 
This is not in line with earlier suggestions (Jin 1990) 
also. That is because unlike earlier case where the spallation reaction 
$^4He$ +$^4He$ was dealt with in isolation, we study this in 
relation to other reactions prevalent in the disk. We find
that $^4He$ could be dissociated much before it can contribute to 
spallation. However, our work supports Ramadurai \& 
Rees' (1985) conjecture that deuterium may be produced in the 
outer regions of the disk provided the disk is at least as thick as $10^3 x_g$.

In the process of performing the simulation we were faced with a
challenge which was never addressed earlier in the literature. The problem
arises because the inflow under consideration is optically thin vertically,
but optically thick horizontally. As a result, photons emitted form a power-law
spectrum. Question naturally arises, whether these power-law photons are 
capable of photo-disintegration. We find that the answer is yes and that the
calculation of usual photo-disintegration gives a lower limit of the changes in the
composition. In the extreme conditions close to the black hole, such
processes are sufficiently effective to produce neutron disks around black holes.

We thank Mr. A. Ray for carefully reading the manuscript.

{}


\begin{thebibliography}{}

\def\ref#1\par{\parshape=2 0in 14.5cm 1cm 13.5cm {#1} \par}
\parskip=0pt
\parindent=0pt

\bibitem[]{}
Arai K., Hashimoto M., 1992,  A\&A 254, 191
\bibitem[]{}
Bisnovatyi-Kogan G., 1998, Observational Evidence for Black Holes
in the Universe, (Ed.) S.K. Chakrabarti, Kluwer Academic Publishers,
Dordrecht, Reidel, p. 1
\bibitem[]{}
Chakrabarti S.K., 1989, ApJ 347, 365
\bibitem[]{}
Chakrabarti S.K., 1996a, Phys. Rep. 266, 229
\bibitem[]{}
Chakrabarti S.K., 1996b, ApJ 464, 664
\bibitem[]{}
Chakrabarti S.K., 1998, Observational Evidence for Black Holes
in the Universe, (Ed.) S.K. Chakrabarti, Kluwer Academic Publishers, Dordrecht, Reidel, p. 19
\bibitem[]{}
Chakrabarti S.K., Molteni D. 1995, MNRAS 272, 80
\bibitem[]{}
Chakrabarti S.K., Titarchuk L. 1995, ApJ 455, 623
\bibitem[]{}
Chakrabarti S.K., Jin L., Arnett W.D., 1987, ApJ  313, 674
\bibitem[]{}
Chakrabarti S.K., Rosner R., Vainshtein S.I., 1994, Nat 368, 434
\bibitem[]{}
Chakrabarti S.K., Titarchuk L.G., Kazanas D., Ebisawa K., 1996,  A\&AS 120, 163 
\bibitem[]{}
Cox J.P., Giuli R.T., 1968, Principles of Stellar Structure, Gordon \& Breach, New York
\bibitem[]{}
Crary D.J., Kouveliotou C., van Paradijs J.,  et al., 1996, ApJ 462, L71
\bibitem[]{}
Cui W., Zhang S.N., Focke W., Swank J.H. 1997, ApJ 484, 383
\bibitem[]{}
Daly R.A., Loeb, A. 1990, ApJ 364, 451
\bibitem[]{} 
Das T.K., Chakrabarti S.K., 1998, ApJ (submitted)
\bibitem[]{}
Ebisawa K., Ogawa M., Aoki T., et al., 1994, PASJ 46, 375
\bibitem[]{}
Ebisawa K., Titarchuk L., Chakrabarti S.K., 1996, PASJ 48, 59
\bibitem[]{}
Gilfanov M., Churazov E., Sunyaev R.A., 1997, Accretion Disks -- New Aspects, (Eds.) 
Meyer-Hofmeister E., Spruit H., Springer Verlag, Heidelberg, p. 45
\bibitem[]{}
Grove J.E., Johnson W.N., Kroeger R.A., et al., 1998, ApJ 500, 899
\bibitem[]{}
Hashimoto M., Eriguchi Y., Arai K., M\"uller E., 1993, A\&A 268, 131
\bibitem[]{}
Hogan C.J., Applegate J.H., 1987, Nat 390, 236
\bibitem[]{} 
Jin L., 1990, ApJ 356, 501 
\bibitem[]{}
Ling J., Wheaton W.A., Wallyn P., et al., 1997, ApJ 484, 375
\bibitem[]{}
McClintock J., Horne K., Remillard R.A., 1995, ApJ 442, 358
\bibitem[]{}
Miyamoto S., Kimura K., Kitamoto S., Dotani T., Ebisawa K., 1991, ApJ 383, 784
\bibitem[]{}
Mukhopadhyay B. 1998, Observational Evidence for Black Holes in the Universe,
(Ed.) S.K. Chakrabarti, Kluwer Academic Publishers, Dordrecht, Reidel, p. 105
\bibitem[]{}
Mukhopadhyay B., Chakrabarti S.K., 1998, ApJ (submitted)
\bibitem[]{}
Paczy\'nski B., Wiita P.J., 1980, A\&A 88, 23
\bibitem[]{}
Ramadurai S., Rees M.J., 1985, MNRAS  215, 53p
\bibitem[]{}
Ramos E., Kafatos M., Fruscione A., et al., 1997, ApJ 482, 167
\bibitem[]{}
Rees M.J., 1984, ARAA 22, 471
\bibitem[]{}
Rees M.J., Begelman M.C., Blandford R.D., Phinney  E.S., 1982, Nat 295, 17
\bibitem[]{}
Romani R.W., 1998, A\&AS. 133, 583
\bibitem[]{}
Shakura N.I., Sunyaev R.A., 1973, A\&A 24, 337
\bibitem[]{}
Shapiro S.L., 1973, ApJ 180, 531
\bibitem[]{}
Shvartsman V.F., 1971, Sov. Astron. AJ 15, 377
\bibitem[]{}
Sunyaev R.A., Titarchuk L.G., 1980, A\&A 86, 121
\bibitem[]{} Tanaka Y.,  Lewin W.H.G., 1995, X Ray Binaries,
(Eds.) W.H.G. Lewin, J. van Paradijs., E.P.J. van den Heuvel,  Cambridge University Press, 
England, p. 126
\bibitem[]{} 
Taam R.E., Fryxall B.A. 1985, ApJ 294, 303
\bibitem[]{} 
Titarchuk, L., Mastichiadis A., Kylafis N., 1997, ApJ 487, 834
\bibitem[]{} van den Heuvel E.P.J., 1992, Proc. Internatl. Space year, ESA ISY 3, p. 29 
\bibitem[]{} 
van den Heuvel E.P.J., 1998, High Energy Astronomy and Astrophysics, (Eds) P.C. Agrawal \& 
P.R. Vishwanath, University Press, Hyderabad, p. 10 
\bibitem[]{}
Vargas M., Goldwurm A., Laurent P., et al., 1997, ApJ 476, L23
\bibitem[]{}
Wiita P.J., 1982, Comm. Astrophys. 6, 251
\end{thebibliography}
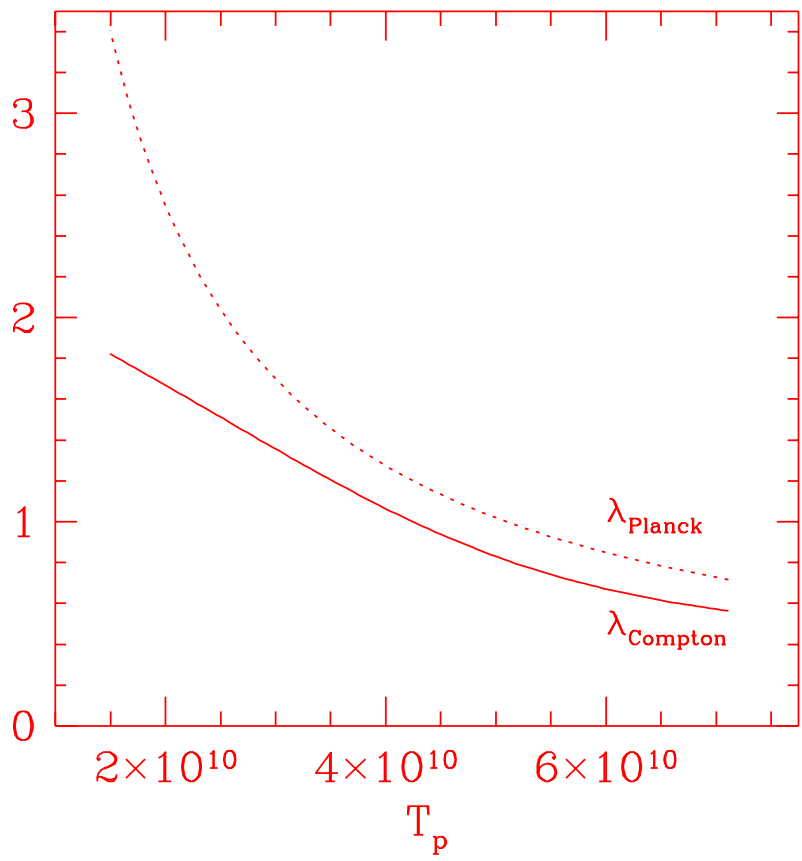
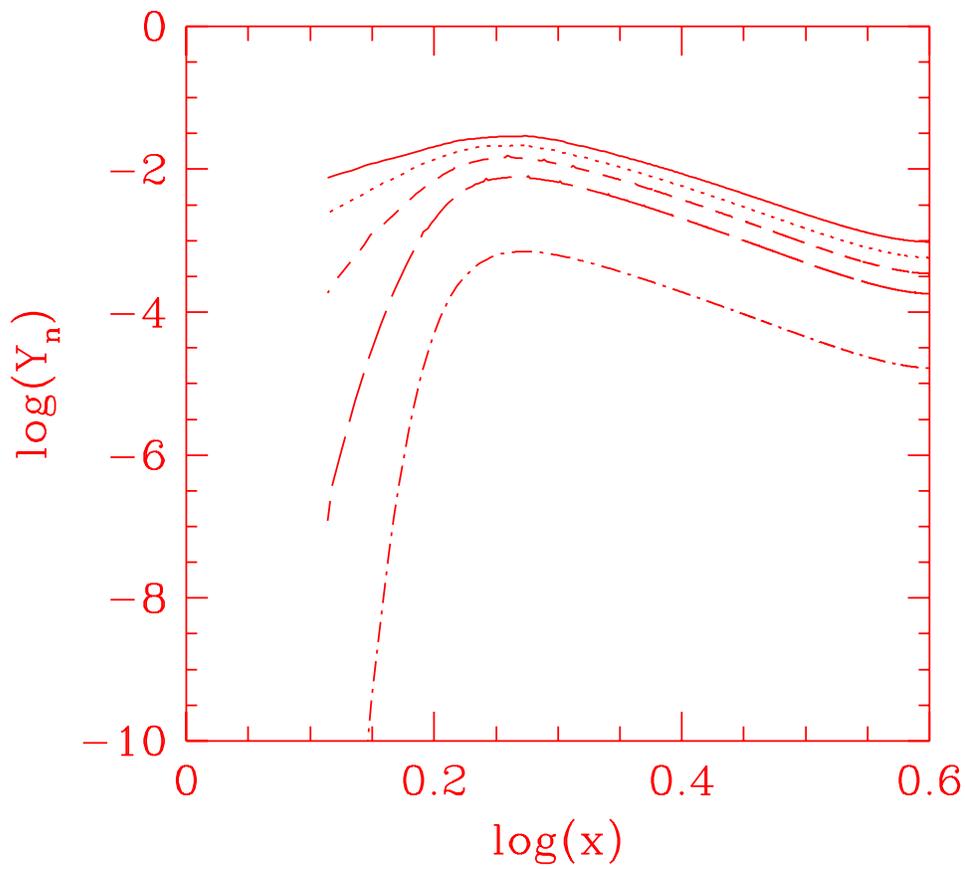
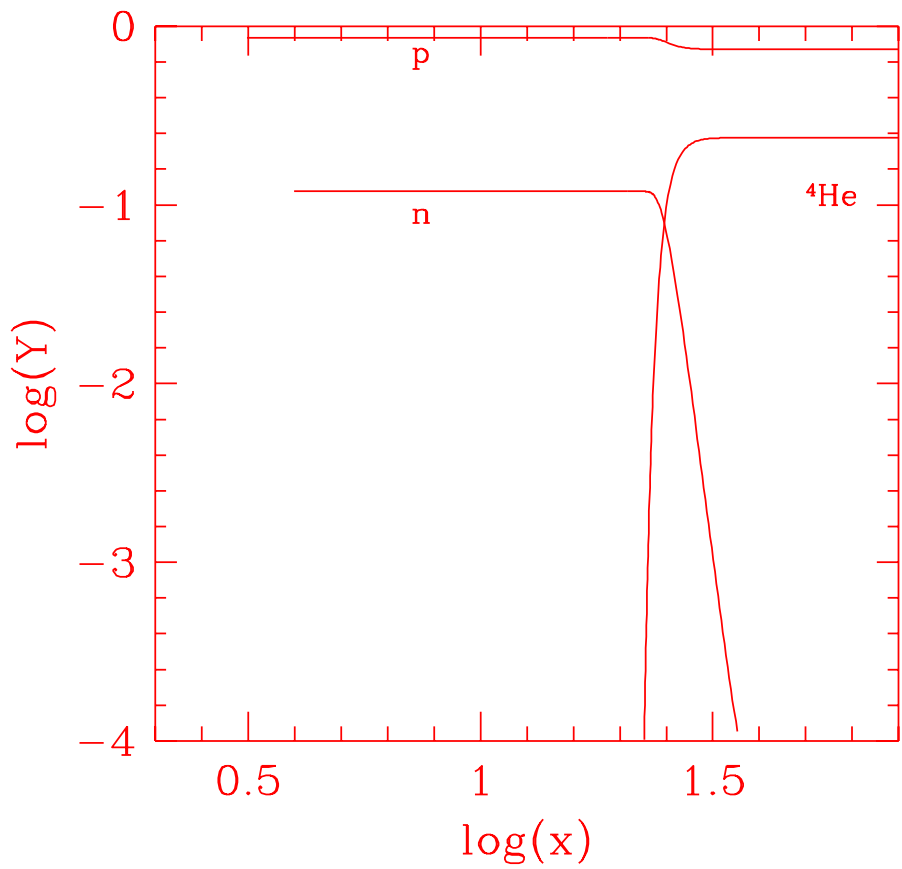
\end{document}